\documentclass[12pt]{article}
\usepackage{amsmath}
\usepackage{amsfonts}
\usepackage{amssymb}
\usepackage{bbm}
\usepackage{verbatim}
\usepackage{dsfont}
\usepackage{booktabs}
\usepackage{slashed}
\usepackage{epsfig}
\usepackage{color}
\usepackage[table]{xcolor}
\usepackage{graphicx}
\usepackage{float}
\usepackage{overpic}
\usepackage{enumerate}
\usepackage{hhline}
\usepackage{multirow}
\usepackage{cite}
\usepackage{xspace}
\usepackage{setspace}

\def\be{\begin{equation}}
\def\ee{\end{equation}}
\def\baed{\begin{aligned}}
\def\eaed{\end{aligned}}
\def\bea{\begin{eqnarray}}
\def\eea{\end{eqnarray}}

\begin{document}
\thispagestyle{empty}

\begin{flushright}
CPHT-RR080.082018\\
\end{flushright}
\vskip .8 cm
\begin{center}
{\Large {\bf  Axions and anomalous $U(1)$'s\footnote{Submitted to the special issue of the International Journal of Modern Physics A ``Field Theory and the Early Universe", in Commemoration of BW2018 and 15 
Years of the SEENET-MTP Network.} }}\\[12pt]

\bigskip
\bigskip 
{
{\bf{ Quentin Bonnefoy}\footnote{E-mail: quentin.bonnefoy@polytechnique.edu}},  
{\bf{Emilian Dudas}\footnote{E-mail: emilian.dudas@polytechnique.edu}}
\bigskip}\\[0pt]
\vspace{0.23cm}
{\it Centre de Physique Th\'eorique, \'Ecole Polytechnique, CNRS, Universit\'e Paris-Saclay, Route de Saclay, 91128 Palaiseau, France}\\[20pt] 
\bigskip
\end{center}

\begin{abstract}
\noindent

Inspired by recent studies of high-scale decay constant or flavorful QCD axions, we review and clarify their existence in effective string models with anomalous $U(1)$ gauge groups. We find that such models, when coupled to charged scalars getting vacuum expectation values, always have one light axion, whose mass can only come from nonperturbative effects. If the main nonperturbative effect is from QCD, then it becomes a Peccei-Quinn axion candidate for solving the strong CP problem.  We then study simple models with universal Green-Schwarz mechanism and only one charged scalar field: in the minimal gaugino condensation case the axion mass is tied to the supersymmetry breaking scale and cannot be light enough, but slightly refined models maintain a massless axion all the way down to the QCD scale. Both kinds of models can be extended to yield intermediate scale axion decay constants. Finally, we gauge flavorful axion models under an anomalous $U(1)$ and discuss the axion couplings which arise.
\end{abstract}

\newpage 
\setcounter{page}{2}
\setcounter{footnote}{0}

{\renewcommand{\baselinestretch}{1.5}


\section{Introduction and Conclusions}

The Peccei-Quinn (PQ) symmetry \cite{Peccei:1977hh} and its light axion \cite{Wilczek:1977pj} (for reviews, see \cite{Ringwald:2014vqa}) is probably the most elegant solution to the strong CP problem. Its implementation in string theory is natural since at tree-level in supergravity there are often continuous PQ like symmetries, usually broken to discrete subgroups by quantum corrections and nonperturbative effects. On the other hand, realistic string models often contain an ``anomalous" abelian gauge symmetry, called $U(1)_X$ in what follows\footnote{We consider the minimal case of one anomalous abelian symmetry, like in the original context it was studied \cite{Nilles:1982ik}, the perturbative heterotic string. Our arguments do however apply to other string models as well, in particular orientifold models, by relabeling appropriately the modulus field, as in our Section \ref{moduli}.}, with anomaly cancellation {\it \`a la} Green-Schwarz (GS). Such a symmetry has multiple phenomenological applications: generating hierarchical fermion masses and mixing angles via the Froggatt-Nielsen mechanism \cite{Ibanez:1994ig}, relating the weak angle to anomaly coefficients \cite{Ibanez:1992fy} and breaking supersymmetry \cite{Binetruy:1996uv}. 

In this note we comment on one additional generic property of models with an anomalous $U(1)$: at the perturbative level, and if there is at least one charged scalar field which gets a vacuum expectation value (vev), such models always contain a potential axion candidate, which can only get a mass by turning on nonperturbative effects (and simultaneously turning on the coupling to gravity in supersymmetric models, where an R-symmetry survives even after inclusion of nonperturbative effects). We study the symmetries responsible for protecting the axion and the conditions under which the axion is light enough for solving the strong CP problem in a heterotic framework with a single charged scalar and hidden sector gaugino condensation, and we conclude that realistic supersymmetry breaking is incompatible with a light enough axion. However, we also give a refined example where nonperturbative dynamics still preserves a massless axion all the way to the QCD scale, even after coupling to gravity. Finally we show that in such a context and irrespective of the details of the model under consideration, gauge invariance fixes completely the couplings of the axion to matter when the charged scalar is used as a flavon field. The couplings to Standard Model (SM) charged fermions are proportional to their anomalous charges and the couplings to the gauge fields to the mixed $U(1)_X G_a^2$ anomalies, where $G_a = SU(3)_c, SU(2)_L, U(1)_Y$ are the SM gauge group factors. Gauge coupling unification conditions alone then determine the ratio of the coupling to the photon to the coupling to the gluons to be $E/N = 8/3$ at the unification scale. These couplings are similar to the ones in the axiflavon/flaxion models \cite{Calibbi:2016hwq}, but the symmetry is now gauged. 

The generic value of the axion decay constant in these simple setups is of order the unification scale. Such values require a tuned or nonstandard cosmology in order to ensure a consistent relic density for the axion. We therefore discuss in the final section models of moduli stabilization which display an intermediate scale axion decay constant. However, aiming for such an intermediate scale decay constant may not be required since several (recent) studies have shown that the resulting cosmology is viable and does not necessarily involve a severe amount of tuning \cite{Steinhardt:1983ia}. Moreover, new proposals for axion dark matter searches are sensitive to GUT scale values for the axion decay constant \cite{Budker:2013hfa}.

\section{Anomalous $U(1)$ models}\label{general} 

\subsection{Perturbative axion in anomalous $U(1)$ models}\label{theorem}
   
 In this section one will prove the following result:\\
 
{\bf Theorem.} In field (string) theory models of a $U(1)_X$ gauge theory with a Stueckelberg (Green-Schwarz) mechanism and at least one charged scalar field acquiring a non-zero vacuum expectation value\footnote{In case of additional $U(1)$ gauge symmetries, anomalous or not, the counting may be different but a similar result always applies.}, at the perturbative level there is always a {\it massless pseudoscalar}.\\

\textbf{Proof.} Let us consider an abelian gauge theory in a Stueckelberg phase, coupled to charged scalars $\Phi_i$ of charges $X_i$, of lagrangian
\be
{\cal L} = | D_{\mu} \Phi_i |^2 - \frac{1}{4} F_{X,\mu \nu}^2 + \frac{1}{2} (\partial_{\mu} a_S + M A_{X,\mu})^2 + \cdots \ , \label{p1}
\ee
where $\cdots$ are other terms like axionic couplings, in which case it is more appropriate to use the term Green-Schwarz rather than Stueckelberg for such a model. Since we are interested in axion-like particles, without loosing generality we only consider in what follows charged scalars having non-vanishing vev's, parametrized as
\be
\Phi_i = \frac{V_i + h_i}{\sqrt{2}} e^{\frac{i \theta_i}{V_i}} \ . \label{p2}
\ee
Gauge transformations act as
\be
\delta A_{X,\mu} = - \frac{1}{g} \partial_{\mu} \alpha \ , \quad \delta \theta_i = X_i V_i \alpha \ , \quad \delta a_S= \frac{M}{g} \alpha  \ . \label{p3}
\ee
From (\ref{p1}) one finds that the Goldstone boson which mixes in the usual way $\partial_{\mu} \theta_X A_X^{\mu}$ with the gauge field is given by (up to a normalization factor)
\be
\theta_X = g X_i V_i \theta_i + M a_S \ . \label{p4}
\ee  
We have therefore $N+1$ potential axions/pseudoscalars, one of which is absorbed by the gauge field via the Higgs mechanism \cite{Binetruy:1998mn}. The perturbative scalar potential is of the form\footnote{It can be checked that the argument below does not change if some of the fields in the scalar potential appear with a complex conjugation.}
\be
V = \sum_{\alpha} \lambda_{\alpha}  \Phi_1^{m_1^{(\alpha)}} \cdots  \Phi_N^{m_N^{(\alpha)}}  + {\rm h.c.} \ ,  \label{p5}
\ee
and gauge invariance imposes the restriction $X_1 m_1^{(\alpha)} + \cdots +X_N m_N^{(\alpha)} = 0$. Simple matrix algebra tells us that the maximal number of independent gauge invariant operators that can be written is equal to $N-1$.  On the other hand, a complete basis of such gauge invariant operators also defines the physical pseudoscalars/axions which can be expressed as a combination of the $\theta_i$'s, since their phases 
\be
\theta_\alpha = \frac{m_1^{(\alpha)} \theta_1}{V_1} + \cdots  +\frac{m_N^{(\alpha)} \theta_N}{V_N}  \label{p6}
\ee 
are automatically orthogonal to the Goldstone boson (\ref{p4}). It is convenient to represent the pseudoscalars above as vectors in a $N+1$ dimensional space, such as for example, up to normalization
\be
\vec{\theta}_X = (g X_1 V_1, \cdots ,g X_N V_N, M) \ , \quad  \vec{\theta}_\alpha = \left(\frac{m_1^{(\alpha)} \theta_1}{V_1},  \cdots  ,\frac{m_N^{(\alpha)} \theta_N}{V_N},0\right)     \ .  \label{p7}
\ee
The scalar potential  (\ref{p5}) then gives masses to at most $N-1$ pseudoscalars. Consequently, there is always (at least) one leftover massless pseudoscalar, which will be a PQ axion candidate if it has the appropriate couplings.  At the perturbative level, it is therefore always possible to define a PQ symmetry in models with an anomalous $U(1)_X$ gauge factor.\\

As one will see in the next sections, nonperturbative effects can generate gauge-invariant potential terms of the form
\be
V_{np} =   \sum_{\beta}  e^{-q_\beta s_0 - i c_{\beta} a_S} \lambda_{\beta}  \Phi_1^{p_1^{(\beta)}} \cdots  \Phi_N^{p_N^{(\beta)}}  + {\rm h.c.} \ ,  \label{p8}
\ee
where $s_0$ is the vev of a scalar and $q_\beta, c_\beta$ are numbers.  Whenever such terms are generated, the leftover massless axion will get a mass from effects other than the usual QCD ones. Such terms can be generated by field-theory nonperturbative effects, instantonic effects in string theory or  quantum gravity effects more generally. 

\subsection{Anomalous $U(1)$: the heterotic case}\label{anomalousheterotic}

In  perturbative heterotic string theory constructions, there is only one possible anomalous $U(1)_X$ and one field, the universal axion-dilaton $S$, transforming non-linearly under gauge transformations. Those act on the different superfields involved as\footnote{We use here the same convention as in \cite{Binetruy:1996uv} to define charges of chiral superfields.}
\begin{eqnarray}
&& \delta V_X \ = \  \Lambda \ + \ {\bar \Lambda} \ , \quad 
\delta \phi^i \ = \ - 2 \ q_i \ \phi^i \ \Lambda \ \equiv \ -2 X^i \Lambda \ , \nonumber \\
&& \delta S \ = \ \delta_{GS} \ \Lambda \ \equiv \ -2 \ X^S \Lambda \ ,   \label{gi1}
\end{eqnarray}
where $X^i,X^S$ define the holomorphic Killing vectors. The modified Kahler potential for the universal axion-dilaton is
\begin{equation}
K \ = \ - \ \ln \ (S + {\bar S} - \delta_{GS} V_X) \label{gi02} 
\end{equation}
and it encodes the Fayet-Iliopoulos (FI) term which appears in the D-term 
\begin{equation}
D_X \ = \ X^I \ \partial_I G \ = \  X^I \ \partial_I K \ = \    q_i \ \phi^i \ \partial_i K \ + 
\ {\delta_{GS} \over 2 (S + {\bar S})} \ , \label{gi01} 
\end{equation}
where in (\ref{gi01}) $G = K + \ln |W|^2$ and we used the gauge invariance of the superpotential $X^I \partial_I W = 0$. We consider $\delta_{GS} > 0$ in what follows. In all known perturbative constructions there always exists in the massless spectrum a field with appropriate sign of the charge (negative in our 
conventions) whose vev is able to cancel perturbatively the (field-dependent) FI term and maintain supersymmetry. We consider the minimal case of one such field, called $\phi$ in
what follows, and normalize its charge to $-1$, following \cite{Binetruy:1996uv}.

Anomaly cancellation conditions relate mixed anomalies $C_i = U(1)_X G_i^2$, where $G_i$ are the various semi-simple factors of the gauge group $G =\prod_{i=1}^N G_i$, such that
\begin{equation}
\delta_{GS} \ = \ {C_1 \over k_1} \ =  \ {C_2 \over k_2} = \cdots= {C_N
\over k_N} \ = \ {1 \over 192 \pi^2} Tr (q_X) \ , \label{gi2}
\end{equation}
where the $k_i$'s are the Kac-Moody levels defining the tree-level gauge kinetic functions
\begin{equation}
f_i \ = \ k_i \ S \ . \label{gi3}
\end{equation}
The last term in (\ref{gi2}) is the FI term, proportional to the mixed $U(1)_X$ - gravitational anomaly, where $Tr (q_X)$ is the sum of $U(1)_X$ charges over all the charged fermions in the spectrum. Therefore, once the FI term is generated, all mixed anomalies have to be different from zero and the theory {\it must} contain charged matter.

\section{A light axion: gaugino condensation and anomalous $U(1)$ in heterotic theories}\label{gauginoCond}

Gaugino condensation in heterotic theories in the presence of the (generic) anomalous $U(1)$ gauge symmetry discussed in Section \ref{anomalousheterotic} has to fulfill the consistency requirements dictated by the coexistence of two local symmetries: supersymmetry and the gauge symmetry. However, although the pure Super-Yang-Mills gaugino condensation superpotential $e^{-3 S/2b_0}$, where $b_0$ is the beta function of the hidden sector, is not gauge invariant, gauge invariance does not forbid gaugino condensation to take place, as was discussed in the heterotic string case some time ago in \cite{Binetruy:1996uv}. It was shown there that the GS cancellation of gauge anomalies restricts the nonperturbative dynamics such that the nonperturbative superpotential is precisely gauge invariant. 

Taking for simplicity a SUSY-QCD model with $N_c$ colors and $N_f < N_c$ flavors and denoting by $Q$ ($\tilde Q$) the hidden sector quarks (antiquarks) of $U(1)_X$ charges $q$ (${\tilde q}$), the GS conditions fix completely the sum of the charges to be
\begin{equation}
C_h \ = \ {1 \over 4 \pi^2} \ N_f (q + {\tilde q}) \ = \ \delta_{GS} k_h \ , \label{gi4}
\end{equation}
where $k_h$ is the Kac-Moody level of the hidden sector gauge group. 
This turns out to be precisely the gauge invariance condition of the
nonperturbative superpotential
\begin{equation}
W_{np} \ = \ (N_c-N_f) \left[ {e^{- 8 \pi^2 k_h S} \over \det (Q {\tilde Q})}
\right]^{1 \over N_c-N_f} \ . \label{gi5}
\end{equation}
Notice that anomaly cancellations (\ref{gi2}) and the structure of the D-term (\ref{gi01}) unambiguously show that the charge 
of the hidden sector mesons $Q {\tilde Q}$ has the same sign as the induced FI term.  Notice also that the charges allow for a perturbative coupling of the form
\be
W_{p} = \lambda_i^{\tilde j} \left(\frac{\phi}{M_P}\right)^{q + \tilde q} Q^i {\tilde Q}_{\tilde j}  \ . \label{gi6}
\ee
Since $\phi$ gets a large vev of the order of the FI term, below the scale of $U(1)_X$ gauge symmetry breaking the perturbative term (\ref{gi6}) becomes a mass term for the hidden sector
quarks and the dynamics of condensation is essentially that of supersymmetric QCD.  
\subsection{The light axion} 

In supersymmetric QCD there is no light axion. The only global anomaly-free symmetry in the UV is an R-symmetry, which is broken explicitly by the mass term. In the model introduced in   \cite{Binetruy:1996uv}
and briefly reviewed above, the mass term is replaced by the coupling  (\ref{gi6}). Then it is easy to check that the following global R-symmetry
\bea
&& \theta' = e^{i \alpha} \theta \ , \quad (Q, {\tilde Q})' (\theta') = e^{\frac{i (N_f-N_c) \alpha}{N_f}}  (Q, {\tilde Q}) ( \theta)  \ , \nonumber \\
&& \phi' (\theta') = e^{\frac{2 i N_c \alpha}{N_f (q + {\tilde q})}}  \phi ( \theta)  \ , \quad S' (\theta') =  S ( \theta) \ , \label{la1}
\eea
is exact and anomaly-free with respect to $SU(N_c)$\footnote{The anomalies with respect to $U(1)_X$ can be canceled by fields from other sectors, for example (MS)SM fields.}. 
It is also spontaneously broken, therefore one expects a massless Goldstone boson. 
More generally, one can combine the R-symmetry above with the gauge symmetry. Indeed:
\bea
&& \phi'(\theta')=e^{iq_\phi\alpha}\phi(\theta) \ , \quad S'(\theta')=S(\theta)-\frac{i}{2}q_S\alpha \ , \nonumber \\
&& Q'(\theta')=e^{iq_Q\alpha}Q(\theta) \ , \quad
\tilde Q'(\theta')=e^{iq_{\tilde Q}\alpha}\tilde Q(\theta)
\label{qb1}
\eea
is a (non-anomalous) R-symmetry of the (non-perturbative) superpotential if:
\be
q_\phi=\frac{2N_c}{(q+\tilde{q})N_f}-\frac{P}{q+\tilde{q}} \ , \quad q_Q + q_{\tilde Q}=\frac{2(N_f-N_c)}{N_f}+P \ , \quad q_S=\frac{N_f P}{4\pi^2k_h} \ ,
\label{qb2}
\ee
where $P$ is a number and $P=0$ corresponds to the R-symmetry (\ref{la1}).

The model has three pseudoscalars, on which we now concentrate. In order to identify the massless axion, it is enough to parametrize the original fields by ignoring any other field than those pseudoscalars. By defining them in order to have canonical kinetic terms, we are led to the parametrization
\be
S =  s_0 \left(1 + i \sqrt{2} \frac{a_S}{M_P}\right) \ , \quad \phi = \frac{V}{\sqrt{2}} e^{\frac{i a_\phi}{V}} \ , \quad M = Q\tilde Q= M_0 I_{N_f \times N_f} e^{i \sqrt{\frac{2}{N_f M_0}} a_M}  \ , \label{la2}
\ee
where $s_0,V$, and $M_0$ are vev's. One combination of those pseudoscalars 
\be
a_X \propto \frac{\delta_{GS}}{\sqrt{2}s_0}a_S+2Va_\phi-(q+\tilde q)\sqrt{2N_fM_0}a_M \label{golds}
\ee
is absorbed by the $U(1)_X$ gauge field. Another one is shifted by the symmetry (\ref{qb1}-\ref{qb2}), which we choose such that it leaves $a_X$ invariant. In the limit where $M_0 \ll V, M_P$, the value of $P$ which achieves this is
\begin{equation}
P=\frac{2N_c}{N_f\left(1+\frac{\delta_{GS}^2M_P^2}{8V^2s_0^2}\right)}+{\cal O} \Big(\frac{M_0}{V,M_P}\Big)  \ , \label{qb3}
\end{equation}
and the associated symmetry current gives us the expression of the physical axion $a_{PQ}$:
\begin{equation}
J_\mu \propto \frac{1}{\frac{1}{V}+\frac{8s_0^2V}{\delta_{GS}^2M_P^2}}\partial_\mu\left(a_\phi-\frac{2\sqrt{2}s_0V}{\delta_{GS}M_P}a_S\right) +{\cal O} \Big[\frac{M_0}{V,M_P}\Big] \equiv \ f_a\partial_\mu a_{PQ} \ , \label{qb4}
\end{equation}
where we identified the axion decay constant
\begin{equation}
\frac{1}{f_a}=\sqrt{\frac{1}{V^2}+\frac{8s_0^2}{\delta_{GS}^2M_P^2}} \ . \label{fa}
\end{equation}
Natural values are of order the unification scale $f_a \sim M_{GUT}$, although smaller values are possible in orientifold models. 

\subsection{Simplified description}

If the scale of hidden sector condensation is well below the scale of $U(1)_X$ gauge symmetry breaking, which we assumed in deriving expressions (\ref{qb3}-\ref{qb4}), there is an approximate decoupling between the hidden sector dynamics and the $U(1)_X$ dynamics.
In particular, in this limit the $SU(N_c)$ dynamics is essentially the one of supersymmetric QCD, which has no light particles, therefore no light composite axion. It should be therefore possible to describe 
accurately the  light axion physics by integrating out the hidden sector. By doing this, one finds an effective superpotential
\be
W_{eff} = W_0 + N_c (\det \lambda)^{\frac{1}{N_c}} M_P^{3-N_f/N_c}\left(\frac{\phi}{M_P}\right)^{\frac{N_f (q + \tilde q)}{N_c}} e^{- \frac{8 \pi^2 k_h S}{N_c}}   \ , \label{la3}
\ee
where the constant $W_0$ was added for the purpose of coupling to gravity later on.  The effect of the hidden sector condensation is therefore of generating a non-perturbative superpotential, sometimes
said to be of ``fractional instanton" type, as compared to ``stringy instanton" effects, which would be proportional to   $e^{- 8 \pi^2 S}$ in our conventions. According to our general discussion in Section \ref{general},
the phase of the nonperturbative term in    (\ref{la3}) defines a physical axion, which is orthogonal to the Goldstone boson $a_X$ precisely when the GS anomaly cancellation conditions (\ref{gi4}) are imposed. One can write explicitly $a_X$ and the (for now) massless axion $a_{PQ}$ by introducing a rotation matrix
\be
a_X = \cos \theta \ a_S + \sin \theta \ a_\phi   \ , \quad a_{PQ} = - \sin \theta \ a_S + \cos \theta \ a_\phi \ ,  \label{la4}
\ee
with $\tan \theta = 2\sqrt{2}s_0 V/(\delta_{GS}M_P)$.  Notice that $a_{PQ}$ coincides with the leading order expression of the axionic current obtained in  (\ref{qb4}).
 
At the global supersymmetry level,  the axion mass is protected by the R-symmetry (\ref{qb3}). However,  after coupling to supergravity, the constant $W_0$ breaks explicitly the R-symmetry and as such the axion will get a scalar potential and therefore a mass \cite{bpr}. Without entering details of moduli stabilization, one expects a scalar potential of the form
\be
\small
 V (a_{PQ}) \sim W_0  N_c (\det \lambda)^{\frac{1}{N_c}} M_P^{3-N_f/N_c}\left(\frac{V}{M_P}\right)^{\frac{N_f (q + \tilde q)}{N_c}} e^{- \frac{8 \pi^2 k_h s_0}{N_c}}  \cos\left(\frac{(q+\tilde q)N_f}{N_c}\frac{a_{PQ}}{f_a}\right) \ ,  \label{la5}
\ee 
where the axion decay constant is given in (\ref{fa}). By using the order of magnitude value for the gravitino mass $m_{3/2} \sim W_0$ and the definition of the IR dynamical scale
\be
\Lambda_L^3 = (\det \lambda)^{\frac{1}{N_c}} M_P^{3-N_f/N_c}\left(\frac{\phi}{M_P}\right)^{\frac{N_f (q + \tilde q)}{N_c}} e^{- \frac{8 \pi^2 k_h S}{N_c}}   \ ,  \label{la7}
\ee 
one finds that this axion can solve the strong CP problem if
\be
m_{3/2} \Lambda_L^3  \ll 10^{-10} f_{\pi}^2 m_{\pi}^2  \ .  \label{la8}
\ee
This is a very strong constraint, which favors in this minimal model low values of the gravitino mass and of the dynamical scale $\Lambda_L$.  Using the fact that in the minimal model of \cite{Binetruy:1996uv} supersymmetry was broken, and $m_{3/2} \sim \Lambda_L^3 / (V M_P)$, one finds, without an additional source of supersymmetry breaking, 
the constraint $m_{3/2} \ll 10^{-14}$ eV, which is not realistic in known mediations of supersymmetry breaking. In this model therefore, an additional source of supersymmetry breaking is necessary, whereas for a gravitino mass corresponding to standard mechanisms for supersymmetry breaking, the axion is too heavy to solve the strong CP problem. 

\subsection{More refined analysis}
       
Let us perform a slightly more general analysis, by keeping the hidden sector mesons in the discussion. The hidden mesons are described by chiral (super)fields of charge $q + \tilde{q}$ and
have a Kahler potential, computed along the flat directions of $SU(N_c)$, equal to
\begin{equation}
K = Tr (M^{\dagger} M)^{\frac{1}{2}} \ . \label{mr1}
\end{equation}
The hidden mesons appear in the full superpotential
\begin{equation}
W = W_{np}+W_p=(N_c-N_f) \left[ {e^{- 8 \pi^2 k_h S} \over \det(M)}
\right]^{1 \over N_c-N_f} + \lambda_i^{\tilde j} \left(\frac{\phi}{M_P}\right)^{q + \tilde q} M^i_{\tilde j} \label{mr2}
\end{equation}
and add a pseudoscalar axionic degree of freedom $a_M$, that is encoded in the parametrization (\ref{la2}). Notice that solving for $M$ in (\ref{mr2}) gives back (\ref{la3}). Out of the original three pseudoscalars, one is the Goldstone boson absorbed by the gauge field 
(\ref{golds}) and the other two are physical, called $a_1$ and $a_2$ in what follows. They can be parametrized by the gauge invariant operators in (\ref{mr2}) and, up to normalization, can be written as
\be
a_1 \sim 8 \sqrt{2} \pi^2 k_h s_0 a_S + \sqrt{\frac{2N_f}{M_0}} a_M \ , \quad a_2 \sim \frac{q+ \tilde{q}}{V} a_\phi + \sqrt{\frac{2}{N_fM_0}} a_M  \ . \label{mr3}
\ee
Notice that they are both orthogonal to the Goldstone boson $a_X$, as enforced by gauge invariance. They are not orthogonal to each other, fact to be taken into account in what follows.
The hidden sector nonperturbative dynamics is giving a mass to the linear combination
\begin{equation}
a_h \sim \frac{a_1}{N_c-N_f} + a_2 \ , \label{mr4}
\end{equation}
whereas the orthogonal combination $a_l$ defined by $(a_h, a_l) = 0$ is the massless (at the global supersymmetric level) axion. For general vev's its expression is relatively involved. However, in the limit we considered in the previous sections $M_0 \ll V,M_P$, one can easily find that
\begin{equation}
a_l \sim a_1 - N_f a_2 \rightarrow a_{PQ}  \label{mr5}
\end{equation}
is precisely the light axion (\ref{qb4},\ref{la4}), obtained by integrating-out from the start the hidden sector mesons in the simplified description. 
\section{A massless axion: the 3-2 model}\label{sec32}

The main problem with the previous minimal model is that the hidden sector nonperturbative dynamics was giving a mass to the axion through supergravity interactions. Nonperturbative dynamics is however often instrumental for stabilizing moduli, in  our case the very modulus involved in the GS mechanism. The natural next step is to identify models in which the hidden sector nonperturbative dynamics leaves an exactly massless axion, even after coupling to (super)gravity. One way to achieve this goes as follows: at the perturbative level, as we proved in Section \ref{theorem}, there is always a massless axion in models with anomalous $U(1)_X$. Suppose now that the hidden sector producing the nonperturbative dynamics has an R-symmetry itself, in the limit in  which the anomalous abelian gauge dynamics is turned off. Then if the condensation breaks spontaneously the R-symmetry, there is another R-axion coming from the hidden sector. In total there are therefore two axions in the limit where gravity is decoupled. By turning on gravity and adding a constant which breaks explicitly the R-symmetry, one (linear combination) of the two axions becomes massive. But the other one remains massless down to the QCD scale and behaves as an ideal candidate for a PQ QCD axion.  Essentially the nonperturbative dynamics is not adding a potential for the axion, but is just stabilizing the GS modulus.  

One explicit model of this type uses for the hidden sector the 3-2 model of supersymmetry breaking \cite{ads}. The gauge group of the model is $G = G_h \times U(1)_X \times \cdots$, where $G_h = SU(3) \times SU(2)$ is the hidden sector gauge group. The nonabelian factor $SU(3)$ is confining with a dynamical scale $\Lambda_3$.  The matter content in the UV contains the chiral multiplets
\bea
&& Q^{\alpha}_i (3,2) \ , \quad L^{\alpha} (1,2) \ , \nonumber \\ 
&& {\bar U}^{i} ({\bar 3},1)   \ , \quad {\bar D}^{i} ({\bar 3},1) \ \rightarrow   \quad {\bar Q}_{\alpha}^i = ( {\bar D}^{i},  {\bar U}^{i})  \   , \label{st1}
\eea
in a self-explanatory notation (notice that the $\alpha$ index of $\bar Q$ is not gauged under $SU(2)$ and only represents a convenient repackaging). The model has two anomaly-free global symmetries, one acting like hypercharge and an R-symmetry:
\bea
&& U(1)_Y \ : \qquad Y(Q) = \frac{1}{6} \ , \quad  Y({\bar U}) = - \frac{2}{3} \ , \quad Y({\bar D}) =  \frac{1}{3} \ , \quad   Y(L) = - \frac{1}{2} \ , \nonumber \\ 
&&U(1)_R  \ : \qquad R(Q) = - 1  \ , \quad   R({\bar U}) = R({\bar D}) = 0  \ , \quad R(L) =   3   \  . \label{st2}
\eea
Below the scale of $SU(3)$ condensation, the dynamics is governed by the gauge invariant operators
\be
X_1 = Q {\bar D} L \ , \quad X_2 = Q {\bar U} L \ , \quad X_3 = \det   {\bar Q}_{\alpha}   Q^{\beta}   \  . \label{st3}
\ee
The low-energy superpotential, compatible with the symmetries and the condensation dynamics, is given by
\be
W_{\rm eff} = \lambda X_1 + \frac{2 \Lambda_3^7}{X_3} \ . \label{st4}
\ee  
The analysis of the potential, including the D-term contributions, shows that $\langle X_1 \rangle $ and $\langle X_3 \rangle $ are non-vanishing whereas $\langle X_2 \rangle $ vanishes.
There are then two pseudoscalars in the hidden sector, the potential axions in the phases of $X_1$ and $X_3$. One linear combination of them will get a mass from the nonperturbative dynamics, and the second one gets a mass from couplings to (super)gravity, as in the model described in the preceding section. 

If we now couple this model to an anomalous $U(1)_X$, we would get an additional pseudoscalar from the high-energy anomalous $U(1)_X$ sector. There is therefore one leftover axion which is massless all the way down to the QCD scale, being a good candidate for a PQ axion. To restrict the superpotential, one could use the anomalous gauge symmetry instead of imposing the hypercharge global symmetry as above. We can for instance give the following charges to the multiplets (where $n$ is some number):
\be
U(1)_X :\
\begin{cases}
X(Q) = \frac{1}{6}+n\\
X({\bar U}) = -\frac{1}{3}\\
X({\bar D}) =  \frac{1}{3}\\
X(L) = -\frac{1}{2}-\frac{n}{3}
\end{cases}
\implies
\begin{cases}
X(X_1) = \frac{2n}{3}\\
X(X_2) = \frac{2(n-1)}{3}\\
X(X_3) = \frac{1}{3}+2n\\
X(\Lambda_3) = \frac{1}{21}+\frac{2n}{7}\\
\end{cases}, \label{st5}
\ee
where, as in the model discussed previously, the condensation scale $\Lambda_3=e^{\frac{-8\pi^2k_3S}{7}}$ is not-gauge invariant anymore due to the $U(1)_XSU(3)^2$ anomaly:
\be
U(1)_XSU(3)^2: C_3=\frac{1}{4\pi^2}\times\left(\frac{1}{3}+2n\right) \ , \quad U(1)_XSU(2)^2: C_2=\frac{1}{4\pi^2}\times\frac{8n}{3} \ ,
\ee
while the nonperturbative superpotential is:
\be
W_{\rm eff} = \lambda \left(\frac{\phi}{M_P}\right)^{\frac{2n}{3}}X_1 + \frac{2 \Lambda_3^7}{X_3}   \  . \label{st6}
\ee
The first term in (\ref{st6}) is a perturbatively generated operator if we assume that $n$ is a multiple of $\frac{3}{2}$. If $\Lambda_3 \ll V$, analogously to the model in the previous  section this axion is essentially a combination of $a_S$ and $a_\phi$.  The axion decay constant will be determined as before and is therefore naturally of the order of the unification scale.

\section{Gauged flavor symmetry and axion couplings to matter}

We now identify the $U(1)_X$ discussed in the preceding sections with a flavor symmetry \cite{Calibbi:2016hwq,Wilczek:1982rv}, since those are naturally anomalous due to the structure of fermion masses and  lead to the GS mechanism \cite{Dudas:1995yu}. Doing this, we will see that we generate axionic couplings for the light physical axion of the theory. Since the explicit examples discussed so far were supersymmetric, we focus on the Minimal Supersymmetric Standard Model (MSSM) in what follows.

We then charge the different MSSM superfields such that the Yukawa terms, as well as the $\mu$-term, now explicitly involve $\phi$:
\be
\baed
W_\text{MSSM} =& \lambda_{u,ij}\Big(\frac{\phi}{M_P}\Big)^{X_{q_i}+X_{u_j}+X_{h_u}}Q_iU_jH_u+\lambda_{d,ij}\Big(\frac{\phi}{M_P}\Big)^{X_{q_i}+X_{d_j}+X_{h_d}}Q_iD_jH_d+\\
&\lambda_{e,ij}\Big(\frac{\phi}{M_P}\Big)^{X_{l_i}+X_{e_j}+X_{h_d}}L_iE_jH_d+\mu\Big(\frac{\phi}{M_P}\Big)^{X_{h_u}+X_{h_d}}H_uH_d  \  . \label{FNMSSM}
\eaed
\ee
A clever choice of $U(1)_X$ charges for the MSSM fields then allows to account for, or at least soften, the mass hierarchies and the $\mu-$problem of the MSSM. We note that the $U(1)_X$ charge of $\phi$ makes it possible to choose most, if all, of the MSSM charges to be positive, consistently with the GS conditions (\ref{gi2}).

Starting from this superpotential, one can work out the couplings of the physical axion to the MSSM fields. Triangle loop diagrams combined with the GS term give for instance the coupling of the axion to QCD gauge fields:
\be
{\cal L} \supset \frac{\sum_i(2X_{q_i}+X_{u_i}+X_{d_i})}{64\pi^2}\frac{a_{PQ}}{f_a}Tr(G\tilde G)= \frac{C_3}{16}\frac{a_{PQ}}{f_a}Tr(G\tilde G) \ ,
\ee
where $a_{PQ}$ is given by the expression in (\ref{qb4})\footnote{This assumes that the axion is mostly carried by $a_\phi$ and $a_S$, which requires that every other dynamics breaking the PQ symmetry in the hidden sector or in the MSSM happens at a much lower energy.}, its decay constant $f_a$ in (\ref{fa}) and $C_3$ is the $SU(3)$ gauge anomaly coefficient which appears in (\ref{gi2}). Note that the domain wall number $N_\text{DW}=\sum_i(2X_{q_i}+X_{u_i}+X_{d_i})$ can be chosen equal to $1$ with a consistent choice of charges for the Higgs doublets. This expression can be understood as a modification of the QCD kinetic function (\ref{gi3}) when the quarks are integrated out:
\be
f_3 = k_3S-\frac{C_3}{2}\ln\left(\frac{\phi}{M_P}\right) \ , \label{gi}
\ee
which displays clearly the two canceling contributions to the $U(1)_XSU(3)^2$ anomaly. Similar expressions hold for the other factors of the MSSM gauge groups.

We can deduce from this an interesting prediction of such models if we embed them in unified theories. Indeed, in such a case the anomaly coefficients are linked at the unification scale. For instance, for $SU(5)$ unification, the MSSM gauge couplings verify $g_3^2=g_2^3=\frac{5}{3}g_Y^2$, while the fact that $S$ determines all the gauge kinetic functions gives $g_Y^2k_Y=g_2^2k_2=g_3^2k_3$ and the GS conditions impose $\frac{C_3}{k_3}=\frac{C_2}{k_2}=\frac{C_Y}{k_Y}$. All this can be combined to get $C_3=C_2=\frac{3}{5}C_1$. Thus, the ratios of the couplings of the axion to the MSSM gauge fields are determined: for instance we get that the ratio (at the GUT scale) between the electromagnetic coupling and the gluons coupling is
\be
\frac{E}{N}=\frac{8}{3} \ . \label{enratio}
\ee
We stress that (\ref{enratio}) is valid not only in flavor models of the type (\ref{FNMSSM}), but in any anomalous $U(1)$ model in which $SU(5)$ unification of gauge 
couplings is imposed. Indeed, (\ref{enratio}) is enforced uniquely by unification and the kinetic function (\ref{gi}), determined by gauge invariance. 

There are also couplings of the axion to the spin of fermions arising from (\ref{FNMSSM}):
\begin{equation}
\frac{\partial_{\mu} a}{f_a} (\overline{\psi_{L,I}}X_{L,I} \gamma^{\mu} \psi_{L,I}+\overline{\psi_{R,I}}X_{R,I} \gamma^{\mu} \psi_{R,I}) \ .
\end{equation}
 Their strength is given by the $U(1)_X$ charges of the MSSM fields, so the lighter generations are more coupled than the heavier ones. Besides, once expressed in terms of mass eigenstates, those couplings can be off-diagonal in flavor space, leading to possible flavor-changing currents \cite{Choi:2017gpf}. However, if the axion dynamics lies at the string/GUT scale, all those effects are very much suppressed and evade current constraints. Still, since the couplings to the first generation of the MSSM are not specifically suppressed, recently proposed experiments \cite{Budker:2013hfa} could have the sensitivity to probe such string scale decay constants in the near future.

 \section{Comments on moduli stabilization and intermediate scale decay constants}\label{moduli}
 
 Moduli stabilization and axions in string models with anomalous $U(1)$ were studied in various papers \cite{ArkaniHamed:1998nu} and the issue of axion mass and decay constant in string theory in various  works, see e.g.  \cite{Svrcek:2006yi,Buchbinder:2014qca}. 
 
 In the context of models of the type discussed in our note, the value of the gravitino mass  is highly correlated to the stabilization of the moduli. One should distinguish the case where the coupling to supergravity lifts the axion mass, like in the model in Section \ref{gauginoCond}, from the case where it does not, like in Section \ref{sec32}. In the first case, there is a strong correlation between the values of the gravitino mass and the axion mass such that keeping the axion light requires very small values of the gravitino mass. It was shown that in minimal models the requirement of  ``uplifting"  the vacuum energy to zero is only compatible with large values of the gravitino mass\cite{Dudas:2005vv}. In more sophisticated models with several charged scalars the gravitino mass can be reduced to the TeV range \cite{Dudas:2007nz}, but still far from the small values needed to keep the axion light enough. In other stabilization schemes, it is still possible to keep the axion light enough with more realistic values of the gravitino mass, see e.g.  \cite{Buchbinder:2014qca}. On the other hand, for models in which coupling to supergravity does not lift the axion mass, like in our Section \ref{sec32}, the scale of supersymmetry breaking is completely decorrelated from the axion mass, which then only gets a mass from QCD nonperturbative effects. 

The moduli stabilization in Sections \ref{gauginoCond} and \ref{sec32} was also enforcing a high-scale axion decay constant due to the $U(1)_X$ D-term expression (\ref{gi01}). This can be relaxed in models where the moduli sector is slightly more complex. For example, let us consider a model of two moduli and a charged superfield:
\be
K=-\frac{3}{2}\ln(T_1+\overline{T_1}-\delta_1V_X)-\frac{3}{2}\ln(T_2+\overline{T_2}+\delta_2V_X)+\phi^\dagger e^{-2V_X}\phi \ ,
\label{2moduli}
\ee
on which the anomalous $U(1)_X$ symmetry acts as follows:
\be
\delta V_X =  \Lambda + {\bar \Lambda} \ , \quad \delta \phi = 2 \phi \Lambda \ , \quad \delta T_1 = \delta_1 \Lambda \ , \quad \delta T_2 = -\delta_2 \Lambda \ .
\ee
The $U(1)_X$ D-term potential $V_D=\frac{g_X^2}{2}(\vert\phi\vert^2+\frac{3\delta_2}{4(T_2+\overline{T_2})}-\frac{3\delta_1}{4(T_1+\overline{T_1})})^2$ now allows for a high scale stabilization of the moduli with a small or intermediate scale $\phi$. To illustrate this, we furthermore assume that there are two hidden strong sectors $1$ and $2$, with gauge kinetic functions given by:
\be
f_1=\frac{T_1}{4\pi} \ , \quad f_2=\frac{n_2T_1+n_1T_2}{4\pi} \ , \quad \text{where } n_i=\pi\delta_i \ \text{are integers}\ ,
\ee
such that the group $1$ is anomalous with respect to $U(1)_X$ whereas $f_2$ is gauge invariant. Strong dynamics can then generate couplings of the type:\footnote{Those nonperturbative effects have periodicity $T_i=T_i+1$ and are called stringy instanton effects. The other option is to use fractional instanton effects, like in Section \ref{gauginoCond}, which would be, with the present section notations, of the type $e^{-2\pi T_i/N}$ where $N \in \mathbb{N}$.}
\be
W=W_0+A\phi^{n_1}e^{-2\pi T_1}+Be^{-2\pi(n_2T_1+n_1T_2)} \ .
\ee
In order to compute the vacuum of the theory, we assume that the uplift of the vacuum energy does not depend on the axions (e.g. {\it \`a la} KKLT \cite{Kachru:2003aw}). Thus, as far as the axions are concerned we look at first order for the supersymmetric vacuum:\footnote{Those three equations can be combined to check that the D-term potential vanishes.}
\be
\baed
&D_\phi W \equiv W_\phi+K_\phi W=An_1\phi^{n_1-1}e^{-2\pi T_1}+\overline{\phi}W= 0\\
&D_{T_1} W=-2\pi A\phi^{n_1}e^{-2\pi T_1}-2\pi n_2Be^{-2\pi(n_2T_1+n_1T_2)}-\frac{3}{2(T_1+\overline{T_1})}W= 0\\
&D_{T_2} W=-2\pi n_1Be^{-2\pi(n_2T_1+n_1T_2)}-\frac{3}{2(T_2+\overline{T_2})}W= 0 \ ,
\eaed
\ee
and we solve this set of equations given the value of $m_{3/2}=\frac{\vert W\vert e^{K/2}}{M_P^2}$ and assuming that $W\approx W_0$ and $\vert\phi\vert^2 \ll T_{1,2}^{-1}$, which we eventually check to be valid:
\be
\frac{T_2+\overline{T_2}}{T_1+\overline{T_1}}=\frac{n_2}{n_1} \ , \ \ 2\pi n_2(T_1+\overline{T_1})e^{-2\pi n_2(T_1+\overline{T_1})}=\frac{3W_0}{2B} \ , \ \ \vert\phi\vert=\left\vert\frac{W_0e^{\pi(T_1+\overline{T_1})}}{n_1A}\right\vert^{\frac{1}{n_1-2}} .
\ee
If we choose for instance $m_{3/2}=10$ GeV, $n_1=3$ and $n_2=1$, we numerically get $T_1+\overline{T_1}=3(T_2+\overline{T_2})\approx 6M_P$ and $\vert\phi\vert\approx10^{11}$ GeV, which implies an intermediate scale for the physical axion. However, in this setup the axion mass is tied to the supersymmetry breaking scale and cannot be light enough to provide a proper QCD axion. To cope with this, one can for instance implement the configuration (\ref{2moduli}) within the 3-2 model of Section \ref{sec32}. This amounts to consider the following superpotential (where all fields are those defined either above or in Section \ref{sec32}):
\be
W=W_0+\lambda\left(\frac{\phi}{M_P}\right)^{\frac{2n}{3}}X_1+\frac{2\Lambda_3^7}{X_3}+Be^{-2\pi k_2(n_2T_1+n_1T_2)} \ , \text{ with } \Lambda_3=e^{\frac{-2\pi k_1 T_1}{7}} \ .
\ee
There is as expected a massless axion in the low-energy limit, and its associated decay constant can be of intermediate scale: for instance, choosing $n_1=n_2=1,n=6,k_1=17, k_2=4$ and $\phi\approx 10^{12}$ GeV, one finds $X_1^{1/3}\approx 10^{12}$ GeV and a gravitino mass of $\approx 10^{-4}$ eV (consistent with gauge mediation of supersymmetry breaking).

\section*{Acknowledgments}

The authors acknowledge partial support from the ANR Black-dS-String, and thank Michele Cicoli and Stefan Pokorski for enlightening discussions.
 

\end{document}